\documentclass[aps,pra,a4paper,twocolumn,groupedaddress,floatfix,showpacs,showkeys]{revtex4}

\usepackage{graphicx}
\usepackage{dcolumn}
\usepackage{bm}

\advance\oddsidemargin by -\textwidth   
\divide\oddsidemargin by 2      
\advance\oddsidemargin by -1in      
\evensidemargin \oddsidemargin

\begin{document}

\title{Bose-Einstein condensation in a simple Microtrap}


\author{S.~Schneider$^1$, A.~Kasper$^1$, Ch.~vom~Hagen$^1$, M.~Bartenstein$^1$, B.~Engeser$^1$, T.~Schumm$^1$}
\author{I.~Bar-Joseph$^2$}
\author{R.~Folman$^1$, L.~Feenstra$^1$}
\email[Contact: ]{Feenstra@physi.uni-heidelberg.de}
\homepage[~\\Website: ]{http://www.bec.uni-hd.de}
\author{J.~Schmiedmayer$^1$}
\affiliation{$^1$Physikalisches Institut, Universit\"at Heidelberg Philosophenweg 12, 69120 Heidelberg, Germany\\
 $^2$Department of Condensed Matter Physics, The Weizmann Institute of Science, Rehovot 76100, Israel}


\begin{abstract}
A Bose-Einstein condensate is created in a simple and robust
miniature Ioffe-Pritchard trap, the so-called Z trap. This trap
follows from the mere combination of a Z-shaped current carrying
wire and a homogeneous bias field. The experimental procedure
allows condensation of typically $3\times 10^5$ $^{87}$Rb atoms
in the $\left|F=2, m_F=2\right>$ state close to any mirroring
surface, irrespective of its structure, thus it is ideally suited
as a source for cold atom physics near surfaces.
\end{abstract}

\maketitle

Bose-Einstein condensates are at the heart of research in quantum
optics. Since the first realization~\cite{Ans95-198,Dav95-3969},
various methods of creating a condensate have been shown, ranging
from large-scale magnetic traps to miniature surface
traps~\cite{Ott01-230401,Haensel2001} as well as optical dipole
traps~\cite{Barrett2001a}. Many  fascinating properties of this
form of matter have been
studied~\cite{kett99-97,NatureColl,GSUBecBib}.

Manipulation of ultra cold atoms using surface mounted
micro-structures, so-called atom chips~\cite{Fol00-4749,fol02},
promises accuracy and versatility in manipulating atoms at the
quantum level, and perhaps even the implementation of quantum
information processing~\cite{BouwmeesterBook}. For a
comprehensive review on surface mounted micro-traps, see
Ref.~\cite{fol02}. One of the main experimental challenges is to
create a simple source of ultra-cold (ground state) atoms for the
experiments, irrespective of the structures on the surface.

In this Rapid Communication we report on the creation of a
sizable ($\sim\!3\times 10^5$~atoms) Bose-Einstein condensate in
a Z-wire trap, which yields a simple, easy to handle and robust
small-scale Ioffe-Pritchard trap~\cite{Haa01-043405}. By virtue
of the trapping wire being part of the atom chip mounting, both
the experimental procedure leading to a condensate, and its
alignment to the atom chip are independent of the actual structure
of the atom chip, or any other reflecting surface. Thus it
provides the means to combine the versatile atom chip assembly and
the stringent UHV demands for Bose-Einstein condensation with a
high number of atoms.


The most basic microtrap uses the magnetic field minimum created
by superposing the magnetic field of a current in a straight wire
with a homogeneous magnetic bias field perpendicular to
it~\cite{Frisch33,fol02}. The two fields add up to a 2D quadrupole
field along the wire, which can guide atoms. The trap depth is
given by the homogeneous bias field, the field gradient is
inversely proportional to the wire current. Such a geometry lends
itself to miniaturization of the wire size using micro-fabrication
techniques. In this way traps can be made even steeper, while
maintaining mechanical stability and robustness. A typical example
is our atom chip~\cite{Fol00-4749}, where the 1~-~200~$\mu$m wide
wires are micro fabricated in a 2~-~5~$\mu$m thick gold layer on a
silicon substrate.

By bending the ends of the wire one creates slightly more complex
structures that provide a three dimensional
confinement~\cite{Den99-291,Haa01-043405,Rei99-3398}.\\
a) The U-trap: by bending a wire in a U-shape, the two fields
from the leads close the guide along the base.
The result is a three dimensional quadrupole field, with a trap minimum $B_{0}=0$.\\
b) The Z-trap: if the leads are pointing in opposite directions
(Z-shaped wire) a field component parallel to the base remains,
giving a Ioffe-Pritchard type trap with a non-zero minimum $B_{0}>0$.

A common property of these traps is the scaling of the strong
linear confinement in the transverse direction, inversly
proportional to the wire current $I_W$ and the distance $z_{0}$
between the wire center and the trap minimum.

\begin{equation}
z_{0} \propto\frac{I_W}{B_{x}}, \hspace{0.7cm}
B'_{\bot}=\frac{\partial B}{\partial x,z}
\propto\frac{{B}_{x}^2}{I_W}. \label{randbo}
\end{equation}
Here $B_{x}$ is the homogeneous bias field perpendicular to the
wire. In the case of a Z-trap, the additional longitudinal field
arising from the leads turns the bottom of the quadrupole trap
into a harmonic potential. The transverse angular trap
frequencies $\omega_{x,z}$ in this potential scale as:
\begin{equation}
\omega_{x,z} \propto \frac{B'_{\bot}}{\sqrt{B_0}}. \label{trapw}
\end{equation}

Such wire traps provide a linear field gradient given by
Eq.~(\ref{randbo}). This enables a large flexibility to first
trap a large number of atoms and second to efficiently compress
the trap to small volumes and high trap frequencies for effective
evaporative cooling. For instance, using a wire current of $I_W =
50$~A and properly tuned bias fields, thermal atoms can be
trapped up to 3~mm from the wire center. By only varying the bias
fields, the trap can be compressed to field gradients in excess
of 400~G/cm at a trap distance of 1.5~mm from the wire center,
reaching trap frequencies of $\omega_{x,z} > 600$~Hz. The large
distance allows a micro-fabricated structure to be placed between
the wire and the trap. This is the principle advantage of the
experimental setup. An additional benefit of such small-scale
circuits is that their low self-inductance enables rapid changing
and switching of the potentials.

\begin{figure}
\includegraphics[width=6.5cm]{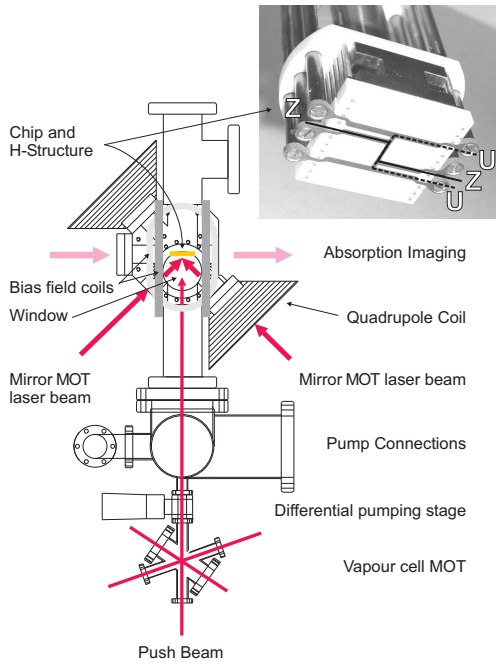}
\caption{\label{zmounting} Scheme of the experimental setup,
showing the vacuum chamber, the quadrupole coils and the
$45^\circ$ laser beams for the mirror-MOT (see text). The inset
shows the copper H-structure in the atom chip mounting (seen from
below, atom chip not mounted). The current paths for the U- and
the Z-trap are indicated. }\end{figure}

As presented in Fig.~\ref{zmounting}, the experimental apparatus
mainly consists of a vertically aligned double magneto-optical
trap (MOT) setup with a mirror
MOT~\cite{Lee96-1177,Pfau96,Rei99-3398,Fol00-4749} in the upper
ultra-high vacuum chamber. The atom chip gold layer serves as the
mirror~\cite{Fol00-4749}. The magnetic quadrupole field for the
MOT is supplied by external coils or by the field of the U-trap
(U-MOT). For the mirror MOT to work, the magnetic axis of the
quadrupole must be tilted by $45^\circ$ with respect to the
mirror. To maintain the optical access to the trap the coils are
wound in a conical shape. Two counter-propagating laser beams
overlap the quadrupole axis, being reflected on the mirror at
$45^\circ$. Two additional counter-propagating laser beams run
horizontally, parallel to the mirror and perpendicular to the
plane of the $45^\circ$-reflected beams. In the case of the
U-MOT, the $45^\circ$ angle of the field axis follows naturally
from the geometry of the wire and the bias field.

The fields for the Z-trap and the U-MOT are derived from currents
through the adequate ports of a monolithic H-shaped copper
structure, see the inset of Fig.~\ref{zmounting}. The central bar
of the H-structure forms the base wire of both the U- and the
Z-trap. The base wire length for the U-MOT is 14.5~mm (outer
ports), while for the Z-trap it is 7.25~mm (inner ports). The
device is tightly fitted in an insulating ceramic, allowing any
structure to be mounted directly against it. The distance between
the center of the base wire and the surface of an atom chip is
1.2~mm. The $1.2\times 1$~mm$^2$ cross-section of the base wire
withstands currents of up to 50~A for over a minute without
significant heating. Thus it enables trapping of large atom
numbers both near and far away from the chip-surface, while
preserving the ultra high vacuum in the upper chamber.

The base of the trap mounting is made from stainless steel tubes
welded to a CF63 UHV-flange, allowing a coolant flow through the
mounting. The flange also carries four high-power current
feedthroughs for operating the H-structure and a 35-pin connector
to access the atom chip. The ceramic block containing the
H-structure is fastened to the stainless steel part of the
mounting by a copper rod, acting as a heat sink. The mounting is
hung from its flange in the upper, ultra high vacuum, chamber of
the setup, which is machined from a single piece and is fitted
with optical quality windows. The distance between the atom trap
center and the closest window is $\sim\!40$~mm.

A regular vapour cell MOT in the lower part of the setup traps
$\sim\!10^9$ $^{87}$Rb atoms from the background gas obtained
from a heated Rb-dispenser ($P\!\sim\!10^{-8}$~mbar). A
continuous laser beam, the push beam, transfers the atoms through
a differential pumping stage and over a 40~cm vertical distance
from the vapour cell MOT chamber($P\!\sim\!10^{-8}$~mbar) to the
UHV
chamber($P\!\sim\!10^{-11}$~mbar)~\cite{BartensteinDiplomarbeit,Wohlleben2001}.

The experimental cycle starts by loading the mirror-MOT by the
push beam for 20~s, resulting in a cloud of $10^9$ atoms located
$\sim\!3$~mm below the atom chip.
Next, the quadrupole field from the mirror-MOT is replaced in 300~ms by that of the U-MOT.
Using a wire current of $I_W=~25.4$~A and bias field of $B_{x}=6.5$~G, a field gradient of 14~G/cm is reached.
The brief transfer to the U-MOT compresses the cloud and moves it to 2~mm from the surface.
During this process we observe no atom losses.

After establishing the U-MOT, all magnetic fields are switched
off for 15~ms to allow optical molasses cooling to
$\sim\!50~\mu$K. It should be noted that the atom chip surface is
not a perfect mirror, due to the structures on the chip. However,
the diffraction from these edges does not harm in any aspect the
functionality of the mirror-MOT, nor does it inhibit molasses
cooling.
A 200~$\mu$s optical pumping pulse to the $\left| F=2,
m_{F}=2\right>$ state prepares the atoms for the transfer to the
Z-trap. Optical pumping increases the number of trapped atoms by a
factor of 3. With careful matching of the trap parameters, more
than $10^{8}$ atoms can be loaded into the magnetic trap.

\begin{figure}
\includegraphics[width=7cm]{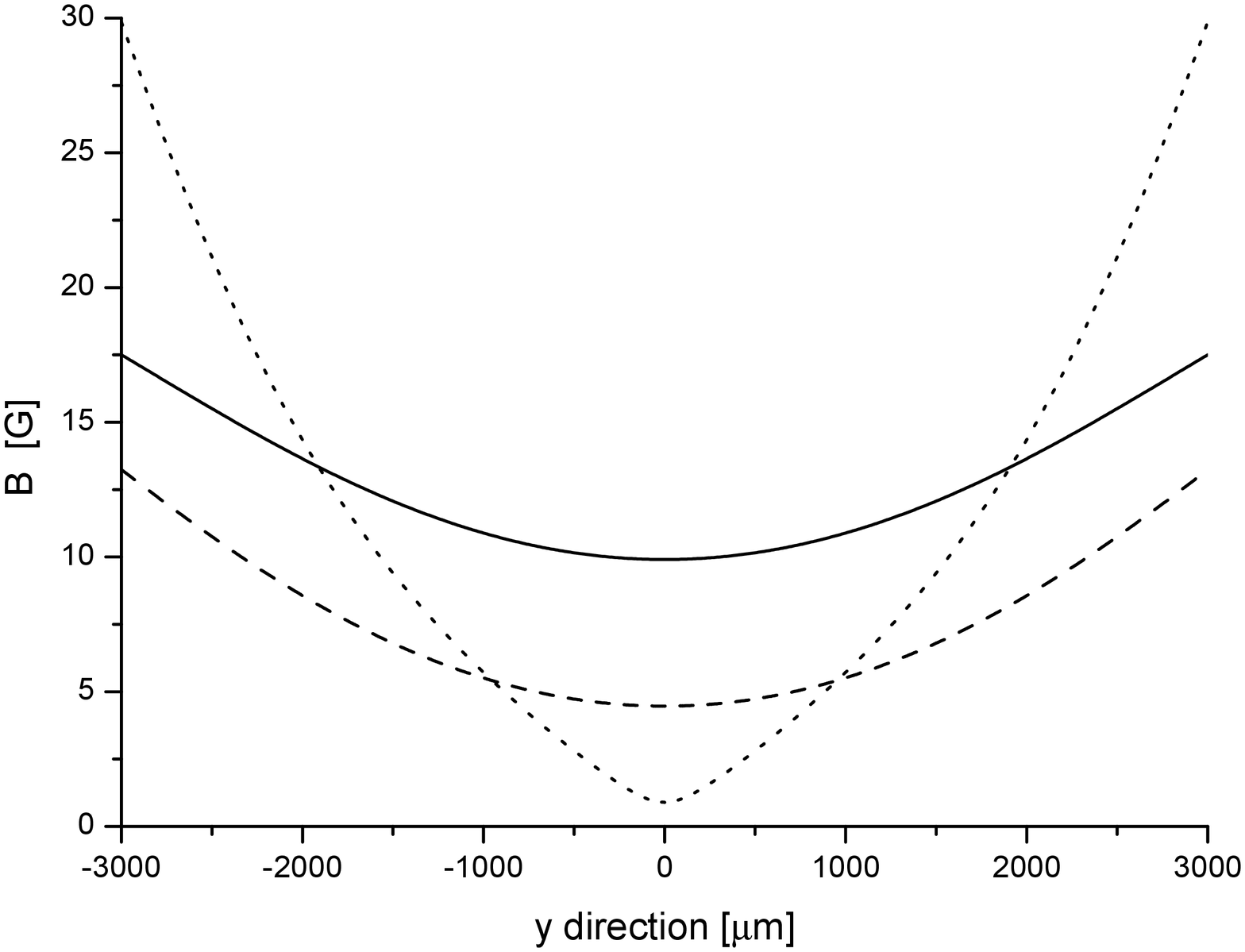}  ~\\

\vspace{0.5cm}
\includegraphics[width=6.5cm]{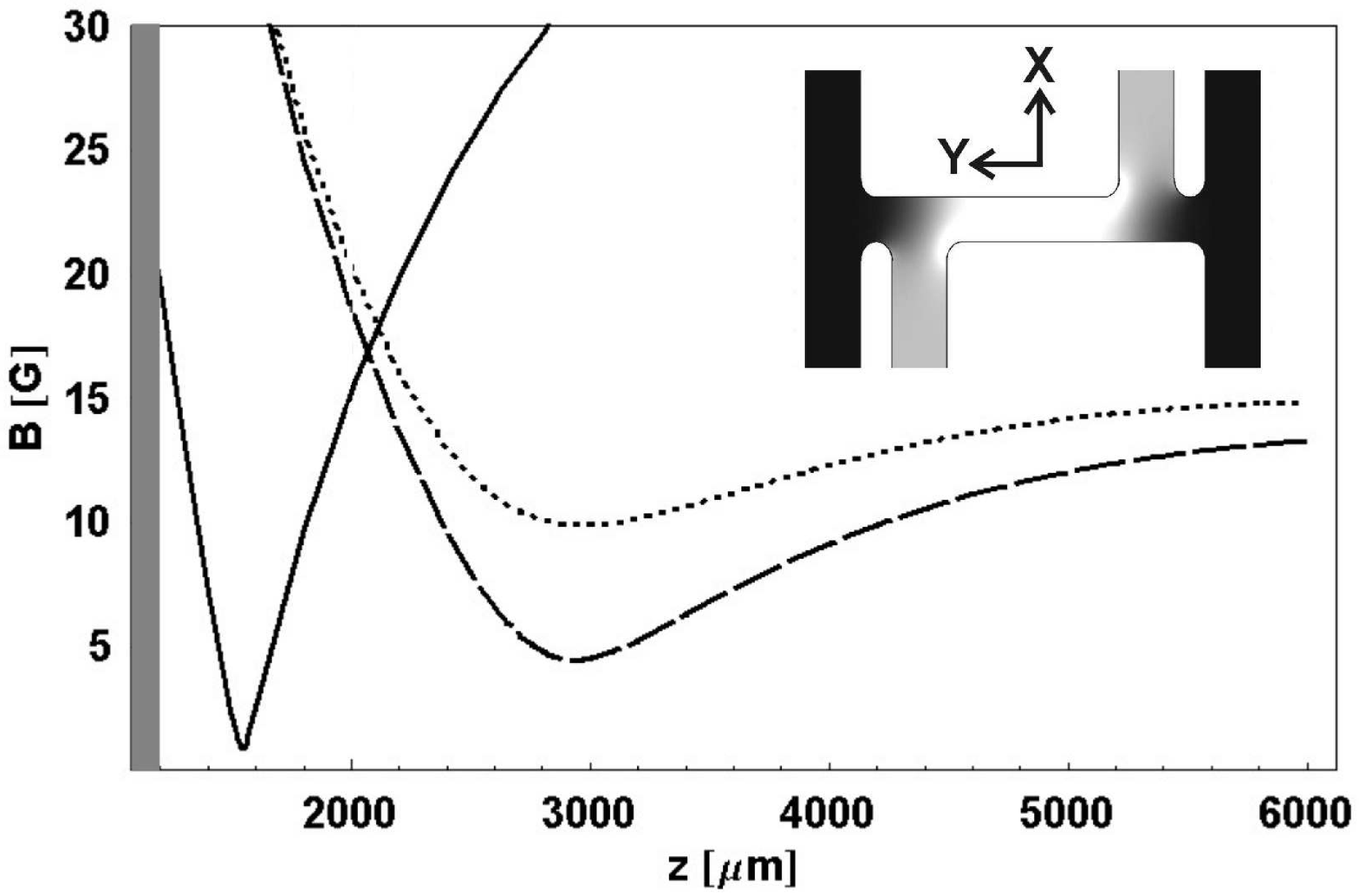}
\caption{\label{bfelder} Potentials of the Z-trap with $I_W=49.7$~A for different bias
field configurations used in the experiment.
Initial trap (dotted line): $B_x=25.4$~G, $B_y=0$; Compressed trap (dashed line):
$B_x=25.4$~G, $B_y=5.5$~G; Final trap (solid line): $B_x=58.3$~G, $B_y=5.5$~G.
On the left axis of the lower panel the chip surface is indicated ($z=1200~\mu$m).
The effect of gravity, along the positive $Z$-direction, is taken into account in
the model.
The inset in the lower panel shows the current density in the H-structure when used for
the Z-trap, a lighter shade corresponds to a higher current density.
}\end{figure}

The trap frequencies for the initial Z-trap, with a Z-current of
49.7~A and a bias field of $B_x = 25.4 $~G (dotted line in
Fig.~\ref{bfelder}), are $\omega_y =2\pi\times 18$~Hz and
$\omega_{x,z} =2\pi\times51$~Hz. Immediately after loading, the
offset field of the Z-trap is reduced by the addition of a bias
field $B_y$, parallel to the base, opposing the fields of the
leads. This compresses the trap transversely and deepens it.
After 100~ms this bias field reaches its final value of
$B_y=5.5$~G (dashed line in Fig.~\ref{bfelder}). The trap is
further compressed in all three directions by linearly increasing
the bias field $B_x$ from 25.4~G to 58.3~G, which also moves the
trap center from initially 1.8~mm to a distance of $300~\mu$m
from the chip surface (solid line in Fig.~\ref{bfelder}). During
the 19~s compression stage, forced evaporative cooling is applied
by RF-radiation from a coil-antenna outside the vacuum chamber.
The frequency is linearly ramped down from 19~MHz to typically
600~kHz. The simultaneous processes of compression and cooling
avoid crushing the initially large atom cloud into the chip
surface.
For the final trapping potential, the angular oscillation
frequencies are $\omega_{x,z} = 2\pi\times 600$~Hz and
$\omega_{y} =2\pi\times 70$~Hz (solid line in Fig.~\ref{bfelder}).
The compression increases the transverse field gradient by more
than a factor of 5, from 80~G/cm to 440~G/cm. The wire current is
kept constant throughout the procedure.

\begin{figure}
\includegraphics[width=6.5cm]{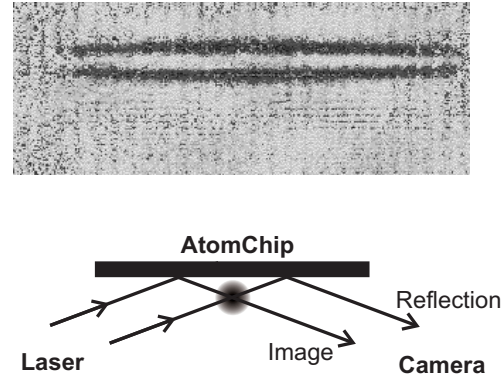}
\caption{\label{ReflectedAbsorption} Absorption image of a
compressed atom cloud ($T\sim5~\mu$K, $N\!\sim\!3\times10^5$)
$\sim$ 35~$\mu$m away from the atom chip surface. The angle of
the imaging laser is adjusted so that the lower image is the
direct shadow of the cloud in the beam reflected off the chip.
The upper image is due to the part of the beam that strikes the
atom cloud before being reflected. The lower panel shows an
exaggerated scheme of the setup (actual angle $\sim\!0.3^\circ$).
}\end{figure}

The atom cloud is studied by absorption imaging with a weak
($120~\mu$W/cm$^2$) circularly polarised laser beam, resonant to
the $\mathrm{F}=2\rightarrow \mathrm{F}'=3$ transition. The
imaging system uses a 8-Bit CCD camera and achromatic optics with
a resolution of $3~\mu$m. The direction of view is parallel to
the plane of the atom chip and perpendicular to the central bar
of the H-structure. The laser beam grazes the chip, enabling
detailed pictures down to very short distances to the chip. Care
has to be taken in aligning the absorption laser beam, because
slight misalignment can cause severe diffraction stripes to
appear in the images.

A helpful feature of the setup is that the distance of the cloud
to the chip surface may be inferred from slightly tilting the
imaging laser beam; one observes two images of the condensate,
where one results from a partially reflected absorption beam (see
Fig.~\ref{ReflectedAbsorption}).

Modeling the trap geometry with a straightforward approach using
infinitely long leads and infinitely thin wires, neglecting the
finite size of the structure, gives a qualitative estimate for the
parameters of the trap. Detailed simulations of the current
distribution in the Z-structure, employing finite element
algorithms, reveal that the effective current path length
$L_\mathrm{eff}$ in the Z-structure is shorter than the
geometrical value, $L_\mathrm{geom}$, defined by the separation of
the leads along the base wire (see inset in Fig.~\ref{bfelder}).
Using an effective baselength of $L_{\mathrm{Eff}}\sim 6.7$~mm,
instead of $L_{\mathrm{Geom}}=7.25$~mm, in the model with
infinitely thin wires, we reach good quantitative agreement with
the experimental results.

\begin{figure}
\includegraphics[width=\columnwidth]{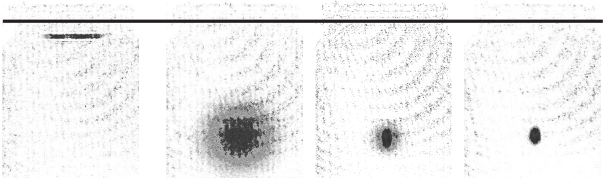}
\caption{\label{exppics} Bose-Einstein condensation is observed in
absorption images taken at different final RF-frequencies
$\nu_f$. From left to right: atoms in the trap, $\nu_f=800$~kHz,
expanded clouds after 15~ms time-of-flight with $\nu_f=800$~kHz
(thermal cloud), $\nu_f=650$~kHz (bimodal cloud) and
$\nu_f=630$~kHz (pure BEC), respectively. Each image size is
$\sim\!1.8\times2.3$~mm. The surface of the atom chip is
indicated at the top of the figure. }\end{figure}

The procedure described above results in a Bose-Einstein
condensate of approx.~$3\times10^5$ atoms, at a density of
$9\times10^{13}$~cm$^{-3}$ and with a transition temperature of
$\sim 600$~nK, see Fig.~\ref{exppics}. The temperature difference
of at least 8 orders of magnitude between the condensate and the
uncooled atom chip can be separated by less than 250~$\mu$m. The
condensate survives for over a second when using RF-shielding at
the final frequency of the cooling ramp.

Since in this work the atom chip is only used as a reflecting
surface for the mirror-MOT, we have shown that a single current
carrying wire in a homogeneous bias field suffices to create and
store a Bose-Einstein condensate near a hot reflective surface,
irrespective of its structure. This makes the setup very well
suited for studying the interaction between degenerate atom
clouds and surfaces.
Since the large volume of the macroscopic Z-trap allows trapping,
storing and cooling of large atom numbers, the technique can
easily be used to load pre-cooled ensembles into small-scale
traps of a different nature, such as miniature optical traps,
traps made of permanent magnets, electric traps, etc., which may
have shallow depths or small trapping volumes, thus lacking the
ability to store sufficient many atoms for compression and
evaporative cooling.

We especially like to thank Toni Sch\"onherr for the fine
mechanical work. This work was supported by the EU project
ACQUIRE, Austrian Science Fund (FWF), and the Deutsche
Forschungsgemeinschaft (Grant Schm1599/2-1). L.F. acknowledges
the support of the Alexander von Humboldt Foundation.

\bibliography{bec_paper}

\end{document}